\documentstyle[preprint,tighten,amssymb,version2,aps]{revtex}
\begin{document}
\draft
\preprint{IFUP-TH 10/93}
\begin{title}
 Renormalization and topological susceptibility on the lattice:\protect\\
 SU$(2)$ Yang-Mills theory.%
\protect\footnote[1]{\ Partially supported by MURST, by a CICYT contract,
and by NATO (grant no.\ CRG~920028).}
\end{title}
\author{Bartomeu~All\'es$^a$, Massimo~Campostrini$^b$,
Adriano~Di~Giacomo$^b$, Yi\u{g}it~G\"{u}nd\"{u}\c{c}$^c$,
and Ettore~Vicari$^b$}
\begin{instit}
$^a$ Departamento de F\'\i sica Te\'orica y del Cosmos,
Universidad de Granada, Spain.

$^b$ Dipartimento di Fisica dell'Universit\`a and I.N.F.N., Pisa, Italy.

$^c$ Hacettepe University Physics Departament, Beytepe Ankara, Turkey.
\end{instit}
\date{January 1993}

\begin{abstract}
The renormalization functions involved in the determination of the
topological susceptibility in the SU$(2)$ lattice gauge theory are
extracted by direct measurements, without relying on perturbation theory.
The determination exploits the phenomenon of critical slowing down to allow
the separation of perturbative and non-perturbative effects. The results
are in good agreement with perturbative computations.
\end{abstract}

\newpage
% ========================= BODY =========================

\narrowtext
\section{Introduction}
\label{Introduction}

The topological susceptibility of the ground state of Yang-Mills
theories is an important parameter for understanding the breaking of the
U$(1)$ axial symmetry.  It is defined by the vacuum expectation value
\begin{equation}
\chi = \int d^4x\left<0|T\left\{q(x)q(0)\right\}|0\right>,
\end{equation}
where
\begin{equation}
q(x) = {g^2\over64\pi^2}\,\varepsilon^{\mu\nu\rho\sigma}
F^a_{\mu\nu}(x)F^a_{\rho\sigma}(x)
\end{equation}
is the topological charge density.  Lattice is the ideal tool to determine
$\chi$ (for a review cf.\ e.g.\ Ref. \cite{DiGiac}).

On the lattice a topological charge density operator with the appropriate
classical continuum limit can be defined as \cite{DiVecchia}
\begin{equation}
q^L(x)\;=\;- {1\over 2^4\times 32 \pi^2}
\sum^{\pm 4}_{\mu\nu\rho\sigma=\pm 1}
\epsilon_{\mu\nu\rho\sigma} {\rm Tr}
\left[ \Pi_{\mu\nu}\Pi_{\rho\sigma}\right]\;\;\;,
\label{Q^L}
\end{equation}
where $\Pi_{\mu\nu}$ is the product of link variables around a plaquette.
Its classical continuum limit is
\begin{equation}
q^L(x)\mathop{\;\longrightarrow\;}_{a\rightarrow 0}
a^4\,q(x)\;+O(a^{6})\;\;\;,
\label{naive}
\end{equation}
where $a$ is the lattice spacing.  We define the lattice bare topological
susceptibility $\chi^L$ from the correlation at zero momentum of two
$q^L(x)$ operators:
\begin{equation}
\chi^L\;=\;\langle \sum_x q^L(x) q^L(0)\rangle\;=
\;{1\over V} \Bigl<\bigl(\sum_x q^L(x)\bigr)^2\Bigr>\;.
\label{chi^L}
\end{equation}
$\chi^L$ is connected to $\chi$ by a non-trivial relationship: the presence
of irrelevant operators of higher dimension in $q^L(x)$ and in the lattice
action induces quantum corrections.  Eq.~(\ref{naive}) must be therefore
corrected by including a renormalization function $Z(\beta)$\cite{Campo1}:
\begin{equation}
q^L(x)\mathop{\;\longrightarrow\;}_{a\rightarrow 0} a^4\,Z(\beta)\,q(x)
\;+O(a^{6})\;\;\;.
\end{equation}
Further contributions arise from contact terms, i.e.\ from the singular
limit $x\rightarrow 0$ of Eq.~(\ref{chi^L}).  They appear as mixings with
the trace of the energy-momentum tensor
\begin{equation}
T(x)\;=\;{\beta(g)\over g} F^a_{\mu\nu}F^a_{\mu\nu}(x)
\label{T}
\end{equation}
and with the unity operator, which are the only available
renormalization-group-invariant operators with dimension equal or lower
than $\chi$.  Therefore the relationship between the lattice and the
continuum topological susceptibility takes the form \cite{first}
\begin{equation}
\chi^L(\beta)\;=\;a^4\,Z^2(\beta)\,\chi\;+\;
a^4\,\tilde A(\beta)\,\langle T(x) \rangle_{\rm N.P.}\;
+\;P(\beta)\langle I \rangle \;+\;O(a^6)\;,
\label{chi_chi^L_0}
\end{equation}
where N.P.\ denotes taking the non-perturbative part.  $T(x)$ is the
correct operator to describe the mixing\cite{Gluon}, since it has no
anomalous dimension.  In order to be consistent with traditional notation,
we will rescale $T(x)$ to the gluon condensate $G_2$, which is defined by
\cite{Shifman1}
\begin{equation}
G_2\;=\;-{1\over4\pi^2b_0} \,\langle T \rangle\;\;\;,
\label{G2e}
\end{equation}
where $b_0=11N/48\pi^2$ is the first coefficient of the $\beta$-function.
In Eq.~(\ref{G2e}) the proportionality constant is fixed by requiring
\begin{equation}
G_2=\left<{g^2\over4\pi^2}\,
F^a_{\mu\nu}F^a_{\mu\nu}(x)\right>\;+\;O(g^4)\;.
\end{equation}
We will then write Eq.~(\ref{chi_chi^L_0}) in the form
\begin{equation}
\chi^L(\beta)\;=\;a^4\,Z^2(\beta)\,\chi\;+\;
a^4\,A(\beta)\,G_2\,
+\;P(\beta)\langle I \rangle \;+\;O(a^6)\;\;\;.
\label{chi_chi^L}
\end{equation}
$Z(\beta)$, $P(\beta)$, and $A(\beta)$ are ultraviolet effects, i.e.\ they
have their origin in the ultraviolet cut-off-dependent modes.  They can be
computed in perturbation theory.  In the present paper, we will show
explicitly that $Z(\beta)$, $P(\beta)$ and $A(\beta)$ (computed
non-perturbatively) are well approximated by the first few perturbative
terms.  This fact is far from trivial, since in this case the perturbative
series are not even Borel-summable (cf.\ Ref.\ \cite{David}).  Moreover, in
principle they could be affected by non-perturbative contributions;
however, there are arguments\cite{Shifman} implying that non-perturbative
effects should not appear in the first few terms of the perturbative
expansion.

In Ref. \cite{ZP} a ``heating'' method was proposed to estimate $Z(\beta)$
and $P(\beta)$ directly from numerical simulations, without
using perturbation theory.
This method has already been employed to study, on the lattice,
the topological properties  of 2-$d$ CP$^{N-1}$ models
\cite{ZP,O3paper,CPNlatt,CPNlatt2}, which are toy models enjoying many
similarities with QCD.  In the present paper we apply the heating method to
the 4-$d$ SU$(2)$ Yang-Mills theory on the lattice, and we show that it can
also be used to disentangle the contribution of the mixing with the
operator $T(x)$.

The main idea of the method is to exploit the fact that renormalizations
are produced by short-ranged quantum fluctuations, at the scale of the
cut-off $a$, whereas physical effects like gluon condensation or
topological properties involve distances of the order of the correlation
length.  When using local updating procedures in Monte Carlo simulations,
fluctuations at distance $l\sim a$ are soon thermalized, whereas
fluctuations at the scale of the correlation length are critically slowed
down when approaching the continuum limit.  For a standard local algorithm,
e.g.\ Metropolis or heat bath, the number of sweeps needed to thermalize
the fluctuations at distance $\xi$ should grow proportionally to $\xi^z$,
with $z\simeq 2$.  Quantities like topological charge, involving changes of
global properties of the configurations, are expected to be even slower to
reach equilibrium, experiencing a more severe form of critical slowing
down.  This is suggested by the fact that in the cooling procedure
\cite{cooling1} the topological charge survives long after the
disappearance of the string tension \cite{cooling3}. In large-$N$ lattice
CP$^{N-1}$ models the autocorrelation time of the topological
susceptibility grows exponentially with respect to the correlation length
\cite{CPNlatt}. A similar phenomenon has been observed in the 2-$d$ U$(1)$
gauge model \cite{U1}.

The heating method consists in constructing on the lattice a smooth
configuration carrying a definite topological charge $Q_0$, and heating it
by a local updating procedure at a given value of $\beta$.  Short-ranged
fluctuations, contributing to $Z(\beta)$ and $P(\beta)$, are rapidly
thermalized, while the initial global topological structure is preserved
for a much larger number of local updatings. This allows us to obtain
estimates of $Z(\beta)$ when heating instanton configurations and of the
mixings when heating flat configurations.

In particular when heating a flat configuration (a configuration with zero
fields, i.e.\ with all the link variables equal to the identity), which has
zero topological charge, we expect the production of instantons at the
scale of $\xi$ to take place much later than the thermalization of local
quantum fluctuations.  The above-mentioned considerations on critical
slowing down effects lead to the expectation that, when $\xi\gg a$, the
heating procedure on a flat configuration should show the following
intermediate stages before reaching full equilibrium:

(a) Short-ranged fluctuations at $l\sim a$ contributing to $P(\beta)$ get
thermalized in a number of updating sweeps independent of $\xi$. The
signal of $\chi^L$ should show a plateau giving $P(\beta)$.

(b) When the number of heating sweeps $n$ increases up to $n\propto\xi^2$,
fluctuations at $l\sim \xi$ start to be thermalized, and gluon condensate
and its mixing with $\chi^L$ sets on. This should produce an increase in
$\chi^L$.

(c) Since a more severe form of critical slowing down is expected to affect
the topological properties, the $\chi^L$ signal should show a second
plateau in which the topological charge is still zero and the whole signal
is mixing to the identity operator and to the trace of the energy-momentum
tensor. This can be checked by cooling back the sample of configurations
and controlling that $Q=0$.

(d) Eventually the modes responsible of the topological structure
will be thermalized and $\chi^L$ will reach its equilibrium value.

Actually, in the SU$(2)$ lattice gauge theory the situation is complicated
by the fact that we are forced to work at small $\xi$ (low $\beta$): the
term involving $\chi$ in Eq.~(\ref{chi_chi^L}) is exponentially suppressed
with respect to the mixing with the identity operator; therefore it becomes
rapidly smaller than the errors and is not detectable at larger $\xi$
(large $\beta$).  Using the Wilson action, the optimal region where to
investigate Eq.~(\ref{chi_chi^L}) is around $\beta=2.5$.  For these values
of $\beta$, the correlation length as obtained from the square root of the
string tension is $\xi_\sigma\simeq 4$, while from the lowest glueball mass
one obtains $\xi_g\simeq 2$.  Of course the relevant correlation length
depends on the quantity we are studying. For example, since the gluon
condensate is closely connected to the glueball propagation, the relevant
correlation length should be $\xi_g$.  In this case of small $\xi$, the
first two regimes (a) and (b) in the heating procedure may become not
clearly distinguishable.  However, we will still have a clear intermediate
plateau where the topological structure is trivial (stage (c)), allowing us
to separate the pure topological contribution from the mixing terms in
Eq.~(\ref{chi_chi^L}).

In Section \ref{numerical} we present our numerical results. They include
data from standard Monte Carlo simulations and from the heating procedure.
In Section \ref{Conclusions} we draw our conclusions.

\section{Numerical results}
\label{numerical}

\subsection{Monte Carlo simulations and perturbative results}
\label{MC}

We performed standard Monte Carlo simulations on a $12^4$ lattice using the
Wilson action and collecting data for $\chi^L$ over an extended range of
$\beta$.  We employed the over-heat-bath updating procedure
\cite{Overheat}.  We also measured the topological susceptibility
$\chi_{\rm cool}$ by using the cooling method \cite{cooling1,cooling2},
which consists in measuring the topological susceptibility on an ensemble
of configurations cooled by locally minimizing the actions (starting from
equilibrium configurations).  The topological content of the cooled
configurations is measured by using $Q^L=\sum_x q^L(x)$.  The topological
susceptibility measured on cooled configurations is seen to gradually reach
a long plateau.  We estimate $\chi_{\rm cool}$ from the plateau
measurements.  Data for $\chi^L$ and $\chi_{\rm cool}$ are reported in
Table \ref{MC-table}.

In Eq.~(\ref{chi_chi^L}), $Z(\beta)$, $A(\beta)$ and $P(\beta)$ can be
calculated in perturbation theory following the field theory prescriptions.
The first few terms of the series are
\begin{equation}
Z(\beta) \;=\;1\,+\,{z_1\over \beta} \,+\,{z_2\over \beta^2}\,+\,...\;,
\label{zetaexp}
\end{equation}
where $z_1$ has been calculated finding $z_1=-2.1448$ \cite{Campo1},
\begin{equation}
A(\beta)\;=\;{b_2\over \beta^2} \,+\,{b_3\over \beta^3}\,+\,...\;,
\label{mixexp}
\end{equation}
where $b_2=1.874 \times 10^{-3}$\cite{b2}, and
\begin{equation}
P(\beta)\;=\;{c_3\over \beta^3}\,+\,{c_4\over \beta^4}\,+\,{c_5\over
\beta^5}\,+\,...\;,
\label{tailexp}
\end{equation}
where $c_3=2.648\times 10^{-4}$ \cite{DiVecchia}
and $c_4=0.700\times 10^{-4}$ \cite{c4}.

While the first calculated terms of $P(\beta)$ fit the data well at large
$\beta$ ($\beta\ge 4$), more terms must be included as $\beta$ decreases.
Lacking an analytical calculation for these terms, one must fit them from
the data. In order to estimate them we, fitted data for $\beta\geq 2.8$,
where the whole non-perturbative signal is smaller than the errors.  One
more term proved to be sufficient for a fit with $\chi^2/{\rm d.o.f}\simeq
1$.  We found
\begin{equation}
c_5\;=\;3.6(4) \times 10^{-4}\;\;\;.
\label{c5}
\end{equation}
We extrapolate this result to get an estimate of $P(\beta)$ at
$\beta\simeq 2.5$.
An overall fit to the $\chi^L$ data of Table \ref{MC-table} gives a value
of $c_5$ consistent with (\ref{c5}), and allows to extract
the signal of dimension 4 in Eq.~(\ref{chi_chi^L}):
\begin{equation}
{Z^2(\beta)\chi\over \Lambda_L^4}\,+\,{A(\beta)G_2\over \Lambda_L^4}
\;\simeq\;0.3 \times 10^5\;\;\;,
\label{dimension4}
\end{equation}
for $\beta\simeq 2.5$.  There is no way within this method to separate the
term proportional to $\chi$ from the mixing to $G_2$ without both a direct
computation of $Z(\beta)$ and $A(\beta)$ and an independent determination
of $\chi$, e.g.\ by cooling.  In fact the term proportional to $G_2$ is
indistinguishable from contributions $O(1/\beta^2)$ (or higher) in
$Z^2(\beta)$.

Evidence for the existence of the mixing to
$G_2$ was obtained by comparing different definitions of $\chi_L$
\cite{b22}. In what follows we will instead obtain the two contributions
independently.

\subsection{Heating an instanton configuration}
\label{zeta}

We start from a configuration $C_0$ which is an approximate minimum of the
lattice action and carries a definite topological charge $Q_0^L$ (typically
$Q_0^L\simeq\pm1$).  We heat it by a local updating procedure in order to
introduce short-ranged fluctuations, taking care to leave the background
topological structure unchanged. We construct ensembles ${\cal C}_n$ of
many independent configurations obtained by heating the starting
configuration $C_0$ for the same number $n$ of updating steps, and average
the topological charge over ${\cal C}_n$ at fixed $n$.  Fluctuations of
length $l\simeq a$ should rapidly thermalize, while the topological
structure of the initial configuration is left unchanged for a long time.
If, for a given $\beta$, we plot $Q^L=\sum_x q^L(x)$ averaged over ${\cal
C}_n$ as a function of $n$, we should observe first a decrease of the
signal, originated by the onset of $Z(\beta)$ during thermalization of the
short-ranged modes, followed by a plateau.  The average of $Q^L$ over
plateau configurations should be approximately equal to $Z(\beta)\,Q_0^L$.
Since we do not expect short-ranged fluctuations to be critically slowed
down, the starting point of the plateau should be independent of $\beta$.

In order to check that heating does not change the background topological
structure of the initial configuration, after a given number $n_c$ of
heating sweeps we cool the configurations (by locally minimizing the
action) and verify that the cooled configurations have topological charge
equal to $Q_0^L$.

We remind that the size of our lattice is $12^4$.  As heating procedure we
used the heat-bath algorithm, which is efficient in updating short-ranged
fluctuations, but is severely affected by critical slowing down for larger
modes.  We constructed on the lattice an instanton configuration according
to the method described in Ref. \cite{Hoek}.  On a $12^4$ lattice we found
that the optimal size of the instanton is $\rho=4$.  We performed also a
few cooling steps in order to smooth over the configuration at the lattice
periodic boundary. After this procedure we end up with a smooth
configuration with $Q^L\simeq 0.90$ (on the lattice $Q^L$ measured on a
discrete approximation to an instanton is exactly 1 only for very large
instantons and in the infinite-volume limit).

In Fig.\ \ref{Zeta-plot} we plot $Q^L({\cal C}_n)/Q_{0}^L$, where
$Q^L({\cal C}_n)$ is the lattice topological charge $Q^L$ averaged over the
ensemble ${\cal C}_n$.  The data in Fig.\ \ref{Zeta-plot} were taken at
$\beta=2.5$ and $\beta=3.0$.  We see clearly a plateau starting from $n=4$
for both values of $\beta$.  The check of the stability of the topological
structure was performed at $n_c=5$.  According to the above-mentioned
considerations, the value of $Q^L/Q_0^L$ at the plateau gives an estimate
of $Z(\beta)$.  We repeated this procedure for other values of $\beta$. The
behavior of $Q^L({\cal C}_n)/Q_{0}^L$ is always very similar to the case
reported in Fig.\
\ref{Zeta-plot}.  The results are presented in Table \ref{Zeta-table}.

A fit of the data to a polynomial
\begin{equation}
Z(\beta)\;=\;1\,-\,{2.1448\over \beta}\,+\,{z_2\over\beta^2}
\label{fitzeta}
\end{equation}
gives $z_2=0.48(4)$ with $\chi^2/{\rm d.o.f.}\simeq 0.3$.

In Ref.\ \cite{first} $z_2$ was estimated by comparing $\chi_{\rm cool}$
with the signal of dimension 4, and it was determined to be $\simeq1.2$.
However, it is clear from Eq.~(\ref{dimension4}) that the estimate included
the mixing with $G_2$.  By using the new determination of $Z(\beta)$, we
can extract $A(\beta)\,G_2$ from the old data; approximating $A(\beta)$
with the first term $b_2/\beta^2$ is consistent with the data and gives
\begin{equation}
{G_2\over\Lambda_L^4} \simeq 0.5(2)\times10^8\;.
\label{G2first}
\end{equation}

\subsection{Heating a flat configuration}
\label{tail}

We now proceed to the analysis of the ensembles ${\cal C}_n$ of
configurations obtained by heating the flat configuration (using the
heat-bath algorithm), for several values of $\beta$.  In Figs.\
\ref{Tail-plot1} and \ref{Tail-plot2} we plot the average value of $\chi^L$
as a function of the number $n$ of heating steps respectively for
$\beta=2.45$ and $\beta=2.5$.  At $n_c=10$ for $\beta=2.45$ and $n_c=10,15$
for $\beta=2.5$ we check by cooling that the topological charge is still
zero.

For both values of $\beta$ we observe long plateaus starting from $n\simeq
10$, which are lower than the equilibrium values of $\chi^L$ (see Table
\ref{MC-table}), but also definitely higher than the estimate of $P(\beta)$
obtained in Section \ref{MC}.  We repeated this procedure for other values
of $\beta$. The behavior of $\chi^L({\cal C}_n)$ is always similar to the
cases plotted in the figures.  Data for the quantity measured on the
plateaus $\chi^L_{\rm pl}$ are reported in Table \ref{Tail-table}.

We believe that $\chi^L_{\rm pl}$ contains the mixings both to the identity
operator and to the trace of the energy-momentum tensor
for the following reasons:

(i) The plateaus observed are long and after cooling back we do not
find  any topological structure.

(ii) The plateau values of $\chi^L$ are systematically higher then the
values of $P(\beta)$ obtained from Eqs.~(\ref{tailexp}) and (\ref{c5}).

(iii) At $\beta\simeq 2.5$, the correlation length relevant to the gluon
condensate is small (it comes from the lowest gluebal mass: $\xi_g\simeq
2$), therefore at $n\sim\xi_g^2$ the fluctuations contributing to the gluon
condensate start to be thermalized; for $n\ge 10$ they could be already
approximately thermalized.

(iv) Data are not inconsistent with a first plateau at $P(\beta)$, followed
by a second one; they are also not inconsistent with the expected shift of
the starting point of the second plateau, when $\beta$ is increased from
2.45 to 2.5.  However, since the change in correlation length is small,
these phenomena can not be detected clearly within our error bars.

The quantity
\begin{equation}
\chi_h(\beta)\;\equiv\;
{\chi^L(\beta)-\chi^L_{\rm pl}(\beta)\over Z^2(\beta)}
\label{chiheat}
\end{equation}
should then measure the physical topological susceptibility.  Data for
$\chi_h$ are reported in Table \ref{MC-table}.  In order to evaluate
$\chi_h(\beta)$, we inserted into Eq.~(\ref{chiheat}) the parametrization
for $Z(\beta)$ given by Eq.~(\ref{fitzeta}), using the fitted value of
$z_2$.
\section{Conclusions}
\label{Conclusions}

In the previous section we obtained two independent estimates of the
topological susceptibility: $\chi_{\rm cool}$, by cooling method, and
$\chi_h$, given by the relationship (\ref{chiheat}).  The comparison is
satisfactory, although $\chi_{\rm cool}$ seems to be systematically lower.
This behavior could be explained by the fact that $Q^L$ underestimates the
topological charge content of the cooled configurations (for the lattice
size we are working with), as we found out explicitly when we constructed
an instanton configuration on the lattice (cf.\ Section
\ref{zeta}).

In the limit $\beta\rightarrow\infty$,
the lattice spacing $a$ is given by the two-loop renormalization formula
\begin{equation}
a\;=\;{1\over \Lambda_L}f(\beta)\;,\;\;\;\;\;\;\;\;\;\;
f(\beta) \;=\;\left( {6\over 11}\pi^2\beta\right)^{51/121}\,
\exp \left( -{3\over 11}\pi^2\beta \right)\;\;\;.
\label{asyformula}
\end{equation}
In Fig. \ref{Asysc-plot} we plot $\chi_h/\Lambda_L^4$ and
$\chi_{\rm cool}/\Lambda_L^4$ versus $\beta$. Scaling is
quite good within the errors.
By fitting to a constant we found:
\begin{equation}
{\chi_h\over \Lambda_L^4}\;=\;3.5(4)\times 10^5 \;\;\;,\;\;\;\;\;\;\;\;\;\;
{\chi_{\rm cool}\over \Lambda_L^4}\;=\;2.8(2)\times 10^5 \;\;\;.
\label{chiovlambda}
\end{equation}

According to the arguments given in Section \ref{tail},
the difference:
\begin{equation}
M(\beta)\;\equiv\;\chi^L_{\rm pl}(\beta)-P(\beta)
\label{mixing}
\end{equation}
should be proportional to the gluon condensate.
Assuming that the first non-trivial term of the perturbative expansion
( Eq.~(\ref{mixexp}) )
gives a good approximation of the function $A(\beta)$, we can estimate
$G_2$ by
\begin{equation}
G_2\;\simeq\;G_M\;\equiv\;{M(\beta)\over b_2/\beta^2}\;\;\;.
\label{Scond}
\end{equation}
In Fig. \ref{Asysc-plot} we plot $G_M/\Lambda_L^4$.  Although the errors
are large, the signal is clear. Fitting the data to a constant we obtained
\begin{equation}
{G_M\over \Lambda_L^4}\;=\; 0.38(6) \times 10^8\;\;\;.
\end{equation}
This result is consistent with Eq.~(\ref{G2first}).  This estimate can also
be compared with an independent determination obtained by studying the
plaquette operator\cite{G2,G22}: $G_2/\Lambda_L^4 = 0.30(2) \times 10^8$
\cite{Gluon}.  The comparison is again satisfactory.

We have shown that Eq.~(\ref{chi_chi^L}) is physically meaningful on the
lattice, since it separates contributions having different physical origin.
Mixings with the unity operator and with the trace of the energy-momentum
tensor are well defined, and we have estimated them.  Estimates of $\chi$
and $G_2$ coming from Eq.~(\ref{chi_chi^L}) and the heating method are
consistent with those obtained independently by other methods.

Finally it is pleasant to notice that our results strongly support the
feasibility of the determination of $G_2$ from the lattice topological
susceptibility in full QCD, i.e.\ in the presence of dynamical quarks, as
proposed in Ref.~\cite{Dynam}.

\subsection*{Acknowledgments}

We would like to thank H.~Panagopoulos for many useful discussions; two of
us (B.~A. and Y.~G.) thank the theory group in Pisa for hospitality.

% ========================= REFERENCES =========================

% ========================= FIGURE CAPTIONS =========================

\figure{Determination of the multiplicative renormalization $Z(\beta)$.
Dashed lines indicate the value of $Z(\beta)$ estimated by averaging data
on the plateau.
\label{Zeta-plot}}

\figure{$\chi^L$ vs.\ the number of updatings when heating a flat
configuration at $\beta=2.45$.  The dashed lines indicate the equilibrium
value of $\chi^L$ and the dot-dashed lines the value of $P(\beta)$ (with
the respective errors).  The solid line shows the value of $\chi^L_{\rm
pl}$ estimated by averaging data on the plateau.  The result obtained by
cooling the configurations after $n_c$ heating sweeps is indicated with
the symbol $\blacksquare$.
\label{Tail-plot1}}

\figure{$\chi^L$ vs.\ the number of updatings when heating a flat
configuration at $\beta=2.5$.  The dashed lines indicate the equilibrium
value of $\chi^L$ and the dot-dashed lines the value of $P(\beta)$ (with
the respective errors).  The solid line shows the value of $\chi^L_{\rm
pl}$ estimated by averaging data on the plateau.  The results obtained by
cooling the configurations after $n_c$ heating sweeps are reported with
the symbol $\blacksquare$.
\label{Tail-plot2}}

\figure{Plot of $\chi_h/\Lambda_L^4$ ( $\Diamond$ ), $\chi_{\rm
cool}/\Lambda_L^4$ ( $\times$ ) and $G_M /\Lambda_L^4$ ( $\blacksquare$ )
vs.\ $\beta$.  Data for $\chi_h$ and $\chi_{\rm cool}$ are slightly
displaced for sake of readibility.
\label{Asysc-plot}}

% ========================= TABLES =========================

\begin{table}
\caption{$\chi^L$, $\chi_{\rm cool}$, obtained by standard Monte Carlo
simulations, and $\chi_h$ vs.\ $\beta$. The column ``stat'' reports the
statistics of the Monte Carlo simulations.}
\label{MC-table}
\begin{tabular}{r@{}lrr@{}lr@{}lr@{}l}
\multicolumn{2}{c}{$\beta$}&
${\rm stat}$&
\multicolumn{2}{c}{$10^5 \chi^L$}&
\multicolumn{2}{c}{$10^5 \chi_{\rm cool}$}&
\multicolumn{2}{c}{$10^5 \chi_h$}
 \\
\tableline
 2&.45   & 80k  & 3&.14(3) & 7&.7(5) & 9&.8(1.8) \\
 2&.475  & 80k  & 2&.91(3) & 6&.5(4) & 7&.6(1.2) \\
 2&.5    & 60k  & 2&.69(4) & 4&.4(3) & 5&.1(1.1) \\
 2&.525  & 80k  & 2&.56(2) & 3&.0(3) & 4&.6(1.0) \\
 2&.7    & 4k   & 1&.84(6) & &&& \\
 2&.8    & 8k   & 1&.60(4) & &&& \\
 2&.9    & 4k   & 1&.36(4) & &&& \\
 3&.0    & 7k   & 1&.21(3) & &&& \\
 3&.25   & 4k   & 0&.93(3) & &&& \\
 3&.5    & 4k   & 0&.69(2) & &&& \\
 3&.75   & 4k   & 0&.58(2) & &&& \\
 4&.0    & 3k   & 0&.45(2) & &&& \\
 4&.5    & 3k   & 0&.33(1) & &&& \\
 5&.0    & 3k   & 0&.227(8)& &&& \\
 6&.0    & 3k   & 0&.127(5)& &&& \\
\end{tabular}
\end{table}

\begin{table}
\caption{Measure of the multiplicative renormalization of $Q^L$,
starting from an instanton of size $\rho=4$ on a lattice $12^4$
and with $Q_0^L=0.90$.
The estimate of $Z(\beta)$ is taken by averaging the data in the range of
$n$ reported in the column ``plateau''.
Since data on the plateau are correlated, as error we report the typical
error of data in the plateau.}
\label{Zeta-table}
\begin{tabular}{r@{}llcc}
\multicolumn{2}{c}{$\beta$}&\multicolumn{1}{c}{ stat }& plateau & $Z^L$ \\
\tableline
 2&.45  &  2k   & 4--7 & 0.20(2) \\
 2&.475 &  3k   & 4--7 & 0.22(2) \\
 2&.5   &  5k   & 4--7 & 0.22(1) \\
 2&.55  &  3k   & 5--7 & 0.23(1) \\
 2&.6   &  2k   & 4--7 & 0.25(2) \\
 2&.8   &  1k   & 5--7 & 0.32(2) \\
 3&.0   &  0.6k & 4--7 & 0.33(2) \\
\end{tabular}
\end{table}

\begin{table}
\label{Tail-table}
\caption{$\chi^L_{\rm pl}$ vs.\ $\beta$ starting from a flat configuration.
$\chi^L_{\rm pl}$ is estimated by averaging
the data in the range of
$n$ reported in the column ``plateau''. As error we report the typical
error of data in the plateau.}
\begin{tabular}{r@{}lrcc}
\multicolumn{2}{c}{$\beta$}& stat & plateau & $10^5 \chi^L_{\rm pl}$ \\
\tableline
 2&.45  &  5k & 10--20 & 2.72(6) \\
 2&.475 &  8k & 10--15 & 2.56(4) \\
 2&.5   & 10k & 11--15 & 2.44(4) \\
 2&.525 & 10k & 10--15 & 2.32(4) \\
\end{tabular}
\end{table}

\end{document}